\documentclass[twocolumn,preprintnumbers,amsmath,amssymb,superscriptaddress]{revtex4}
\usepackage{graphicx}
\usepackage{dcolumn}
\usepackage{bm}
\usepackage{soul}
\usepackage{color}
\usepackage{epstopdf}
\usepackage[version=3]{mhchem}
\usepackage{lipsum}
\usepackage[outercaption]{sidecap}
\usepackage{floatrow}
\usepackage{hyperref}
\usepackage{appendix}

\begin{document}

\title{Theoretical Insights into Non-Arrhenius Behaviors of Thermal Vacancies in Anharmonic Crystals}
\author{Tran Dinh Cuong}
\email{cuong.trandinh@phenikaa-uni.edu.vn}
\affiliation{Faculty of Materials Science and Engineering, Phenikaa University, Hanoi 12116, Vietnam}
\author{Anh D. Phan}
\email{anh.phanduc@phenikaa-uni.edu.vn}
\affiliation{Faculty of Materials Science and Engineering, Phenikaa University, Hanoi 12116, Vietnam}
\affiliation{Phenikaa Institute for Advanced Study (PIAS), Phenikaa University, Hanoi 12116, Vietnam}
\date{\today}

\date{\today}

\begin{abstract}
Vacancies are prevalent point defects in crystals, but their thermal responses are elusive. Herein, we formulate a simple theoretical model to shed light on the vacancy evolution during heating. Vibrational excitations are thoroughly investigated via moment recurrence techniques in quantum statistical mechanics. On that basis, we carry out numerical analyses for Ag, Cu, and Ni with the Sutton-Chen many-body potential. Our results reveal that the well-known Arrhenius law is insufficient to describe the proliferation of vacancies. Specifically, anharmonic effects lead to a strong nonlinearity in the Gibbs energy of vacancy formation.  Our physical picture is well supported by previous simulations and experiments.
\end{abstract}

\maketitle
\section{Introduction}

Vacancies or missing atoms have aroused extensive interest due to their drastic impact on physicochemical properties \cite{1}. For example, one can adopt these point defects to control mechanical behaviors \cite{2}, aging mechanisms \cite{3}, and irradiation responses \cite{4} of metallic systems. Additionally, vacancy engineering can pave the way to design advanced materials for clean energy storage and harvesting \cite{5}.

In that spirit, various experimental approaches have been proposed to enhance the understanding of lattice vacancies \cite{6}. Utilizing differential dilatometry (DD) \cite{7} gives us direct information about the equilibrium vacancy concentration $c_V$. Although DD is widely seen as the most accurate technique, its resolution limit is only $c_V\sim10^{-5}$. Hence, DD measurements are primarily restricted to a narrow temperature region near the melting point $T^m$. This predicament is somewhat improved through positron annihilation spectroscopy (PAS) \cite{8}. The efficient positron trapping in vacancies makes the PAS sensitivity up to $c_V\sim10^{-7}$. Consequently, it is possible to capture the thermal production of vacancies above $\sim0.6T^m$.

Apart from experiments, one often applies density functional theory (DFT) to defective solids \cite{9}. This powerful method can consider all free-energy contributions without adjustable parameters. Unfortunately, there are too many exchange-correlation (xc) functionals to choose from for the internal vacancy surface \cite{10}. Thus, in some circumstances, DFT results exhibit a strong scatter \cite{11}. Furthermore, most DFT studies on vacancies are done at 0 K or within the quasi-harmonic approximation (QHA) \cite{12,13,14}. Developing high-temperature DFT calculations remains a daunting task because of their immense computational cost. 

It is conspicuous that there is a vast temperature gap between experiments and simulations. The basic strategy to bridge the gap is to extrapolate experimental data by the linear Arrhenius equation \cite{15}. Specifically, the vacancy formation enthalpy $H^f$ and entropy $S^f$ are supposed to be unchanged during heating. However, the fitted value of $H^f$ can be appreciably larger than its DFT counterpart \cite{16}. Besides, the generation of vacancies in many metallic substances does not obey the classical Arrhenius ansatz \cite{17}. These anomalous phenomena have raised persistent controversies in the physics community. 

Recently, machine learning (ML) potentials have emerged as a promising tool for modeling crystallographic defects at elevated temperatures \cite{18}. They enable physicists to perform large-scale computations with near-DFT accuracy. Remarkably, some ML studies \cite{19,20} have unveiled the crucial role of intrinsic anharmonicity in breaking down the Arrhenius law. Their findings are qualitatively consistent with classical molecular dynamics (CMD) simulations \cite{21,22}. Notwithstanding, since reference databases are created by DFT, ML potentials still face the xc problem \cite{19,20}. Moreover, a quantitative agreement among ML, CMD, and DFT outputs has not been reached, particularly for the entropic profile \cite{19,20,21,22}. More efforts are needed to adequately address these issues.

Another viable approach for determining the equilibrium vacancy concentration is the statistical moment method (SMM) \cite{23,24,25}. This quantum model is well suited to characterize vibrational excitations in pure metals \cite{26}, solid solutions \cite{27}, and ionic compounds \cite{28}. Interestingly, its mathematical aspects are straightforward and transparent. Therefore, one can complete SMM analyses within a few minutes and obtain valuable information about material properties. For instance, the underlying correlation among melting transition, elastic deformation, and vacancy formation has been elucidated in recent SMM works \cite{24,25}. Nevertheless, non-Arrhenius behaviors of vacancies have not been fully understood. As mentioned above \cite{19,20,21,22}, this limitation is most likely a consequence of combining the SMM with the QHA \cite{23,24,25}.  

Herein, we develop the SMM to achieve a more reliable picture of the vacancy evolution. Theoretical calculations are performed for representative transition metals with the face-centered cubic (FCC) structure, including Ag, Cu, and Ni. Our SMM results are thoroughly compared with existing experiments and simulations.

\section{Theoretical Background}

In the perfect FCC lattice, the \textit{i}th atom is equivalent to an isotropic three-dimensional oscillator with the Hooke constant $k_i$ and nonlinear coefficients $\gamma_{1i}$, $\gamma_{2i}$, $\gamma_i$ \cite{29,30,31}. Applying the Leibfried-Ludwig expansion \cite{32} to the cohesive energy $E_i$ gives us
\begin{eqnarray}
&k_i=\cfrac{1}{2}\left(\cfrac{\partial^2E_i}{\partial u_{i\alpha}^2}\right)_{eq},\quad\gamma_{1i}=\cfrac{1}{48}\left(\cfrac{\partial^4E_i}{\partial u_{i\alpha}^4}\right)_{eq},&\nonumber\\
&\gamma_{2i}=\cfrac{1}{8}\left(\cfrac{\partial^4E_i}{\partial u_{i\alpha}^2\partial u_{i\beta}^2}\right)_{eq},\quad\gamma_i=4\left(\gamma_{1i}+\gamma_{2i}\right),&
\label{eq:1}
\end{eqnarray}
where $u_{i\alpha}$ and $u_{i\beta}$ stand for Cartesian components of the atomic displacement ($\alpha\ne\beta=x,y,$ or $z$). Their average values can be inferred from the force balance criterion as \cite{29,30,31}
\begin{eqnarray}
&\langle u_{i\alpha}\rangle=\langle u_{i\beta}\rangle=\langle u_{i}\rangle=\sqrt{\cfrac{-C_{2i}+\sqrt{C_{2i}^2-4C_{1i}C_{3i}}}{2C_{1i}}},&\nonumber\\
&C_{1i}=3\gamma_{i},&\nonumber\\ &C_{2i}=3k_i\left[1+\cfrac{\gamma_i\theta}{k_i^2}\left(x_i\coth x_i+1\right)\right],&\nonumber\\ &C_{3i}=-\cfrac{2\gamma_i\theta^2}{k_i^2}\left(1+\cfrac{x_i\coth x_i}{2}\right),&
\label{eq:2}
\end{eqnarray}
where $\theta=k_{0B}T$ is the product of the Boltzmann constant $k_{0B}$ and the absolute temperature $T$, $x_i=\hbar\omega_i/2\theta$ is the scaled phonon energy, $\hbar$ is the reduced Planck constant, and $\omega_i$ is the Einstein frequency. Based on the quantum density matrix, the SMM directly associates $\langle u_{i}\rangle$ with quadratic and quaternary moments by \cite{29,30,31}
\begin{eqnarray}
\langle u^2_{i}\rangle=\langle u_{i}\rangle^2+\theta A_{1i}+\frac{\theta}{k_i}\left(x_i\coth x_i-1\right),
\label{eq:3}
\end{eqnarray}
\begin{eqnarray}
&\langle u^4_{i}\rangle=\langle u_{i}\rangle^4+6\theta A_{1i}\langle u_{i}\rangle^2+8\theta^2A_{2i}\langle u_{i}\rangle+3\theta^2A_{1i}^2&\nonumber\\
&+\cfrac{\theta}{k_i}\left(\langle u_{i}\rangle^2+\theta A_{1i}\right)\left(x_i\coth x_i-1\right).&
\label{eq:4}
\end{eqnarray}
Explicit expressions for $A_{1i}$ and $A_{2i}$ were formulated in Ref.\cite{29} via iterative techniques. From these, we can straightforwardly estimate the Helmholtz free energy $F_i$ by \cite{29,30,31}
\begin{eqnarray}
F_i=\frac{1}{2}E_i+F_i^{(QHA)}+F_i^{(ANH)},
\label{eq:5}
\end{eqnarray}
\begin{eqnarray}
F_i^{(QHA)}=3\theta\left[x_i+\ln\left(1-e^{-2x_i}\right)\right],
\label{eq:6}
\end{eqnarray}
\begin{eqnarray}
F_i^{(ANH)}=3\left(\int_0^{\gamma_{1i}}\langle u^4_{i}\rangle d\gamma_{1i}+\int_0^{\gamma_{2i}}\langle u_{i}^2\rangle^2d\gamma_{2i}\right),
\label{eq:7}
\end{eqnarray}
where $F_i^{(QHA)}$ and $F_i^{(ANH)}$ denote quasi-harmonic and anharmonic contributions of atomic vibrations, respectively. Integrating Equation (\ref{eq:7}) provides \cite{29,30,31}
\begin{eqnarray}
&F_i&\approx\cfrac{1}{2}E_i+3\theta\left[x_i+\ln\left(1-e^{-2x_i}\right)\right]\nonumber\\
&+&\cfrac{3\theta^2}{k_i^2}\left[\gamma_{2i}x_i^2\coth^2x_i-\cfrac{2}{3}\gamma_{1i}\left(1+\cfrac{x_i\coth x_i}{2}\right)\right]\nonumber\\
&+&\cfrac{6\theta^3}{k_i^4}\left[\cfrac{4}{3}\gamma_{2i}^2x_i\coth x_i\left(1+\cfrac{x_i\coth x_i}{2}\right)\right.\nonumber\\
&-&\left.2\left(\gamma_{1i}^2+2\gamma_{1i}\gamma_{2i}\right)\left(1+\cfrac{x_i\coth x_i}{2}\right)\left(1+x_i\coth x_i\right)\right].\nonumber\\
\label{eq:8}
\end{eqnarray}

Physically, the appearance of a single vacancy at the \textit{l}th site ($l\ne i$) changes $F_i$ to $F_i^V$ by weakening the metallic cohesion and destroying the cubic symmetry \cite{9}. For simplicity, we assume that $F_i$ and $F_i^V$ have the same analytical form. All structural characteristics of the imperfect system are encoded in coupling parameters $k_i^V$, $\gamma_{1i}^V$, and $\gamma_{2i}^V$ via the Leibfried-Ludwig theory \cite{32} (see Electronic Supplementary Information). Thus, the Gibbs energy of vacancy formation $G^f$ is written by
\begin{eqnarray}
G^f=\sum_{i\ne l}^{N}\left(F_i^V-F_i\right)+P\Omega^f,
\label{eq:9}
\end{eqnarray}
where $N$ is the total lattice-site number, $P$ is the hydrostatic pressure, and $\Omega^f$ is the vacancy formation volume. At $P=0$ GPa, $G^f$ is explicitly expressed in the high-temperature limit $(x_i\coth x_i\approx x_i^V\coth x_i^V\approx1)$ by
\begin{eqnarray}
G^f&=&\sum_{i\ne l}^N\left[\cfrac{1}{2}\left(E_i^V-E_i\right)+\cfrac{3}{2}\theta\ln\left(\cfrac{k_i^V}{k_i}\right)+3\theta^2\cfrac{\gamma_{2i}^V-\gamma_{1i}^V}{k_i^{V2}}\right.\nonumber\\
&-&3\theta^2\cfrac{\gamma_{2i}-\gamma_{1i}}{k_i^2}+12\theta^3\cfrac{\gamma_{2i}^{V2}-3\gamma_{1i}^{V2}-6\gamma_{1i}^V\gamma_{2i}^V}{k_i^{V4}}\nonumber\\
&-&\left.12\theta^3\cfrac{\gamma_{2i}^2-3\gamma_{1i}^2-6\gamma_{1i}\gamma_{2i}}{k_i^4}\right],
\label{eq:10}
\end{eqnarray}
where $E_i^V$ is the defective counterpart of $E_i$. Equation (\ref{eq:10}) helps us deduce $S^f$ and $H^f$ from the following thermodynamic relations
\begin{eqnarray}
S^f=-\frac{dG^f}{dT},
\label{eq:11}
\end{eqnarray}
\begin{eqnarray}
H^f=G^f+TS^f.
\label{eq:12}
\end{eqnarray}
On that basis, the equilibrium vacancy concentration $c_V$ in the dilute limit is determined by
\begin{eqnarray}
c_V=\exp\left(-\frac{G^f}{\theta}\right)=\exp\left(\frac{S^f}{k_{0B}}\right)\exp\left(-\frac{H^f}{k_{0B}T}\right).
\label{eq:13}
\end{eqnarray}

\section{Results and Discussion}
\subsection{Interatomic Potential}

\begin{figure}[htp]
\includegraphics[width=9 cm]{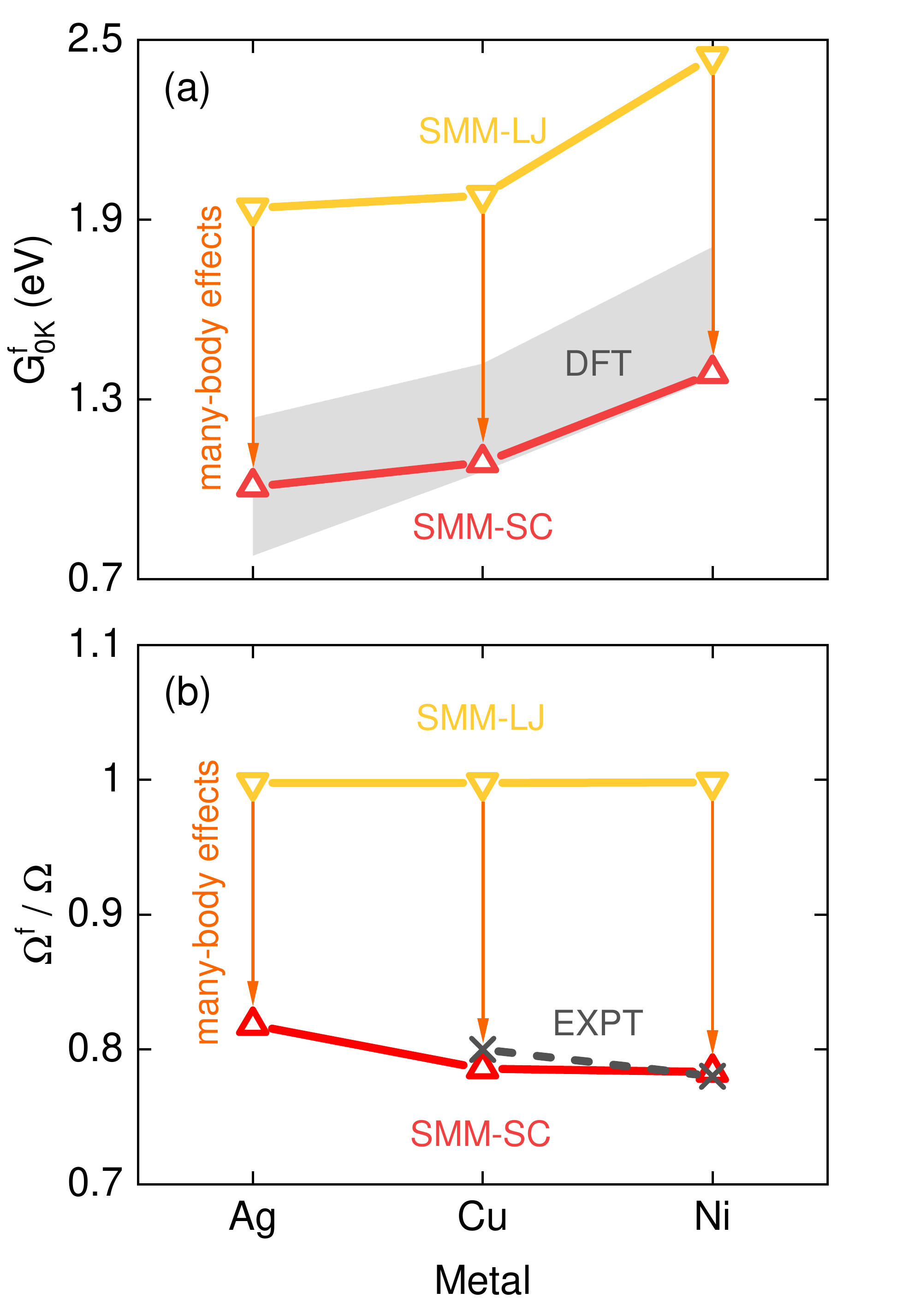}
\caption{\label{fig:1}(Color online) (a) Our SMM calculations for the Gibbs energy of vacancy formation in Ag, Cu, and Ni at zero temperature (open symbols). The shaded area represents DFT data collected by Xing \textit{et al.} \cite{37} and Gong \textit{et al.} \cite{38}. (b) The ratio of the vacancy formation volume $\Omega^f$ to the atomic volume $\Omega$ given by our SMM analyses (open symbols) and previous experiments \cite{15} (cross symbols).}
\end{figure}

In prior SMM works \cite{23,24,25}, the cohesive energy of FCC metals was predominantly modeled by the Lennard-Jones (LJ) pairwise potential as \cite{33} 
\begin{eqnarray}
E_i=\frac{\xi m}{n-m}\sum_{j\ne i}^{N}\left(\frac{\sigma}{r_{ij}}\right)^n-\frac{\xi n}{n-m}\sum_{j\ne i}^{N}\left(\frac{\sigma}{r_{ij}}\right)^m,
\label{eq:14}
\end{eqnarray}
where $r_{ij}$ is the interatomic distance, $\sigma$ is the equilibrium value of $r_{ij}$, $\xi$ is the dissociation energy, and $m$ and $n$ are adjustable parameters. This treatment allows predicting thermodynamic and mechanical properties at breakneck speed. A complete SMM-LJ run merely takes ten seconds on a regular computer. Nonetheless, the lack of many-body interactions can cause an alarming mismatch between the SMM-LJ and other approaches. For example, the Gibbs energy of vacancy formation is overestimated by at least 35 $\%$ in the ground state (see Figure \ref{fig:1}(a)). In addition, the SMM-LJ fails to explain structural relaxation processes around vacancies (see Figure \ref{fig:1}(b)). To overcome these shortcomings, we expand Equation (\ref{eq:14}) via the embedded atom method of Sutton and Chen (SC) \cite{34}, which is
\begin{eqnarray}
E_i=\varepsilon\sum_{j\ne i}^{N}\left(\frac{a}{r_{ij}}\right)^n-2\varepsilon c\sqrt{\sum_{j\ne i}^{N}\left(\frac{a}{r_{ij}}\right)^m}.
\label{eq:15}
\end{eqnarray}
Herein, SC semi-empirical parameters $\varepsilon$, $a$, and $c$ for Ag are taken from Ref.\cite{35}, whereas those for Cu and Ni are extracted from Ref.\cite{36}.

Our choice of the SC potential relies on two principal reasons. Firstly, it is conspicuous that LJ and SC formulas are relatively similar. Hence, the SMM computational efficiency is almost preserved. Secondly, the SMM-SC model can yield reliable predictions about the physical behaviors of vacancies. Indeed, SMM-SC outputs for $G^f_{0\mathrm{K}}$ are entirely in the cold DFT region \cite{37,38}. Besides, the atomic rearrangement in the defective cell is fully captured. As illustrated in Figure \ref{fig:1}, there is excellent accordance between SMM-SC and measured results for $\Omega^f$ \cite{15}.

\subsection{Vacancy Formation at Zero Pressure}

\begin{figure}[htp]
\includegraphics[width=9 cm]{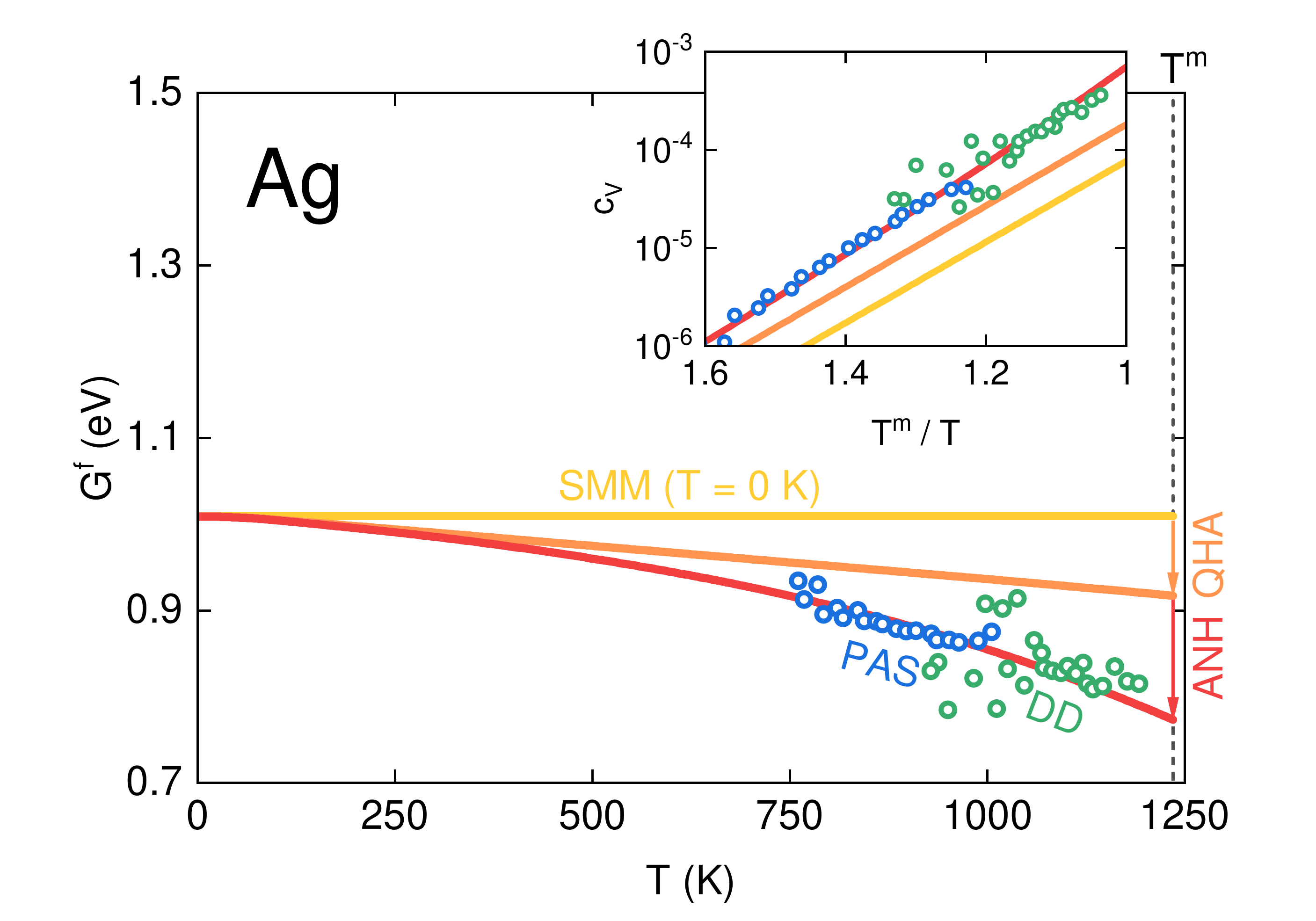}
\caption{\label{fig:2}(Color online) Correlations between the absolute temperature and the vacancy formation free enthalpy of Ag derived from our SMM approximations (thick solid lines) and DD/PAS experiments \cite{39} (open symbols). Inset: The equilibrium vacancy concentration of Ag as a function of the inverse temperature.}
\end{figure}

Figure \ref{fig:2} shows how the vacancy formation Gibbs energy of Ag depends on temperature. Overall, there is a continuous reduction in $G^f$ with increasing $T$. In the QHA, $G^f$ drops linearly from 1.00 to 0.92 eV between 0 and 1235 K. This thermal response agrees qualitatively well with previous SMM studies \cite{23,24,25} on Ag vacancies. Unfortunately, SMM-QHA estimations deviate dramatically from DD and PAS measurements \cite{39}. To go beyond the QHA, we add high-order anharmonic terms to $G^f$ via the thermodynamic integration. Surprisingly, the $G^f$-$T$ profile bends downward and passes through most experimental points \cite{39}. At $T^m$, a significant decline of 0.15 eV is recorded in $G^f$ owing to the inclusion of anharmonicity. From Equation (\ref{eq:13}), one can readily realize that the critical vacancy concentration $c_V(T^m)$ quadruples from $1.8\times10^{-4}$ to $7.2\times10^{-4}$. In other words, anharmonic effects actively promote the proliferation of vacancies in Ag. Our microscopic picture is in consonance with recent ML and CMD simulations \cite{19,20,21,22}.

\begin{figure}[htp]
\includegraphics[width=9 cm]{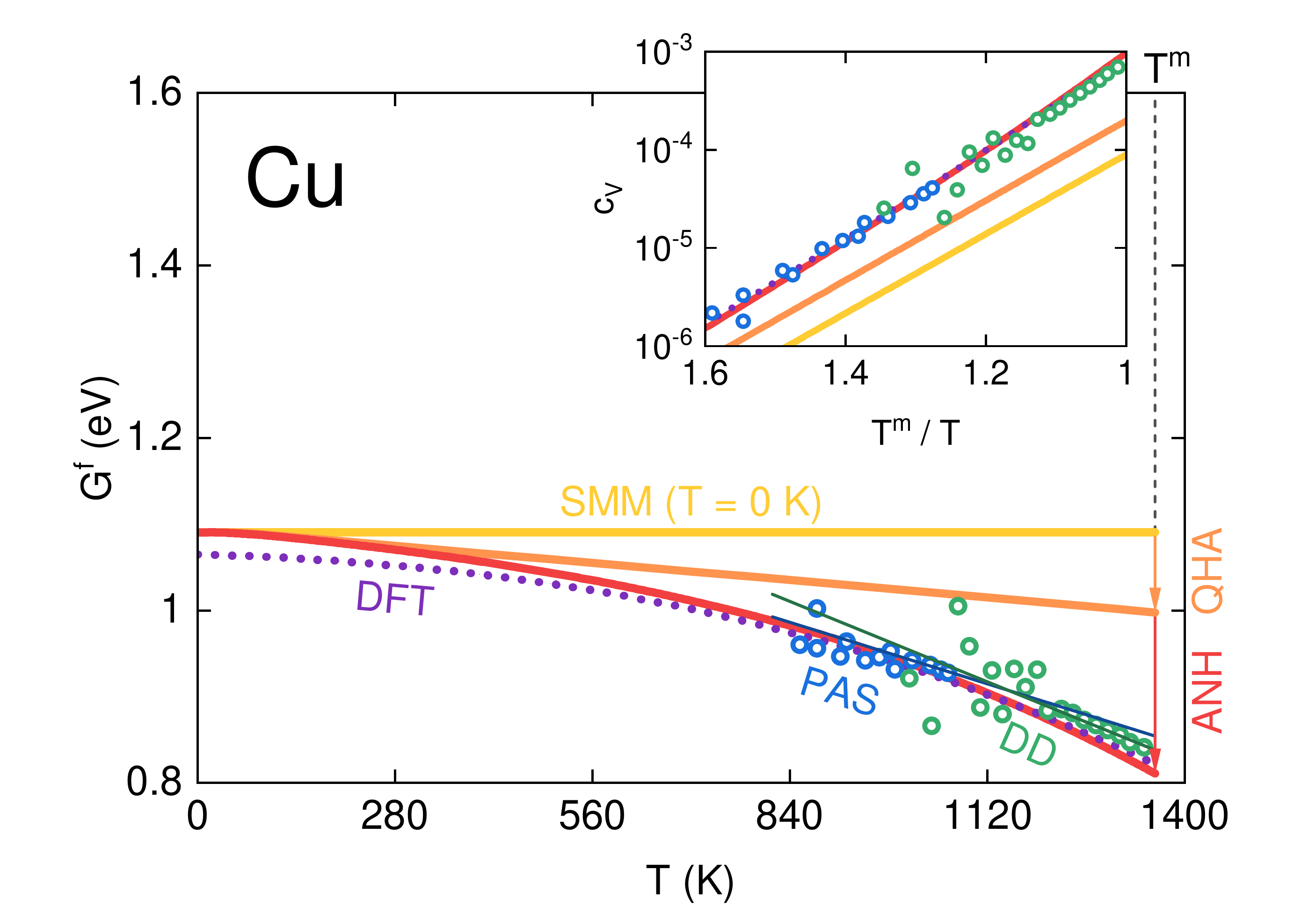}
\caption{\label{fig:3}(Color online) Effects of temperature on the formation Gibbs energy of the Cu vacancy obtained from our SMM calculations (thick solid lines), DFT simulations \cite{42} (dotted line), Arrhenius analyses (thin solid lines), and DD/PAS experiments \cite{39} (open symbols). Inset: The inverse temperature dependence of the equilibrium vacancy concentration in Cu.}
\end{figure}

Figure \ref{fig:3} shows how the absolute temperature influences the Gibbs energy of vacancy formation in Cu. It is worth noting that $G^f$ exhibits a strong nonlinearity during heating. Two distinct Arrhenius functions are required to fit DD/PAS data \cite{39} in the experimentally accessible region. Numerous attempts have been made to decipher this strange phenomenon over the past three decades. According to Neumann \textit{et al.} \cite{40}, the occurrence of divacancies is responsible for non-Arrhenius behaviors. Their empirical analyses have suggested that the divacancy concentration can be up to $0.34c_V$ at the melting point. In contrast, based on CMD, Sandberg and Grimvall \cite{41} have observed the thermal variation of $H^f$ and $S^f$ because of local anharmonic excitations. These vibrational mechanisms alone are sufficient to interpret the curvature in the Arrhenius plot.

Notably, in recent work, Glensk \textit{et al.} \cite{42} have developed the \textit{upsampled thermodynamic integration using Langevin dynamics} (UP-TILD) to resolve long-standing controversies about Cu. Their state-of-the-art approach allows taking into account finite-temperature effects at the DFT level. Having the UP-TILD at hand, Glensk \textit{et al.} \cite{42} have rigorously demonstrated that only 0.1 $\%$ of missing atoms at $T^m$ are divacancies. This negligible proportion has ruled out the Neumann hypothesis \cite{40}. Thus, the anharmonicity is an exclusive source for the violation of the Arrhenius law. UP-TILD simulations with the Perdew-Burke-Ernzerhof (PBE) xc functional \cite{43} of Glensk \textit{et al.} \cite{42} are in good accordance with available DD and PAS measurements \cite{39}. 

As presented in Figure \ref{fig:3}, we can accurately reproduce UP-TILD outputs \cite{42} for $G^f$. The difference between our method and the UP-TILD \cite{42} is less than 0.03 eV in a wide range of 0 to 1358 K. A better consensus will be achieved if the xc energy is treated by the hybrid scheme of Heyd, Scuseria, and Ernzerhof (HSE) \cite{44}. This strategy helps improve DFT-PBE predictions for filled \textit{d}-band noble metals by including nonlocal interactions \cite{37}. In the case of Cu, DFT-HSE computations \cite{37} provide $G^f_{0\mathrm{K}}=1.09$ eV, which completely coincides with the SMM value. Consequently, the SMM can serve as an efficient and reliable tool to investigate the defect evolution in anharmonic crystals.

\begin{figure}[htp]
\includegraphics[width=9 cm]{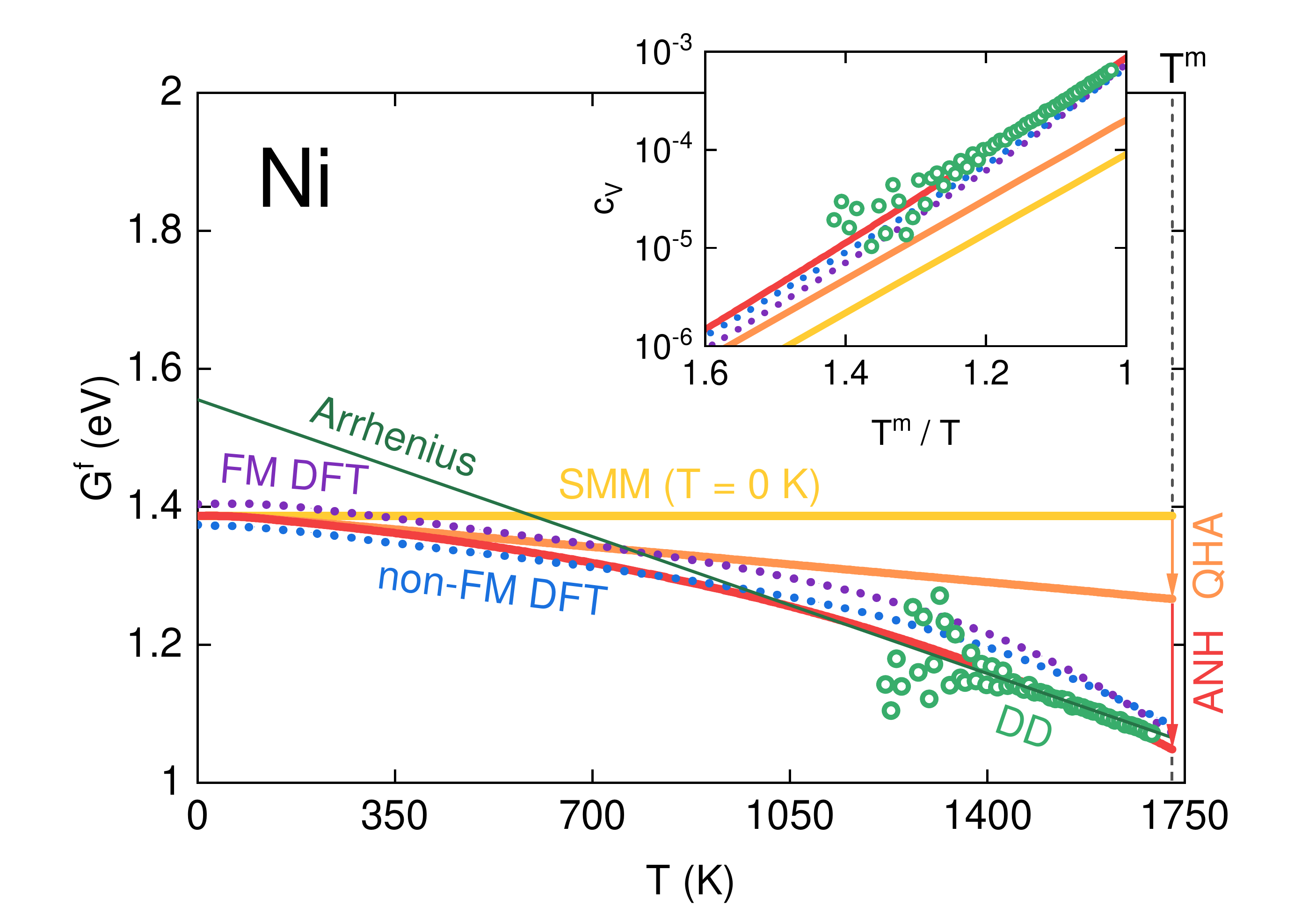}
\caption{\label{fig:4}(Color online) The vacancy formation Gibbs energy of Ni versus the absolute temperature inferred from our SMM calculations (thick solid lines), DFT simulations \cite{38} (dotted lines), Arrhenius analyses (thin solid line), and DD experiments \cite{45} (open symbols). Inset: Impacts of the inverse temperature on the equilibrium vacancy concentration of Ni.}
\end{figure}

Figure \ref{fig:4} shows the temperature dependence of the formation free enthalpy of the Ni vacancy. Experimentally, the well-known Arrhenius law should not be employed to compare practical measurements with quantum-mechanical calculations. This classical extrapolation overestimates $G^f$ at the low-temperature regime. From our perspective, the core problem lies in the nonlinear fluctuation of Ni atoms. Our anharmonic theory suggests that the linear Arrhenius function needs to be replaced with a third-order polynomial as 
\begin{eqnarray}
G^f=g_0+g_1T+g_2T^2+g_3T^3,
\label{eq:16}
\end{eqnarray}
where $g_0$, $g_1$, $g_2$, and $g_3$ are fitting parameters. Indeed, applying Equation (\ref{eq:16}) to existing DD data \cite{45} gives $g_0=1.37$ eV. This figure is close to $G^f_{0\mathrm{K}}=1.38$ eV derived from the SMM and $G^f_{0\mathrm{K}}=1.40$ eV deduced from the \textit{two-stage upsampled thermodynamic integration using
Langevin dynamics} (TU-TILD) \cite{38}. 

Computationally, one can see that SMM results are slightly lower than their TU-TILD counterparts \cite{38}. This discrepancy principally originates from the presence of ferromagnetic (FM) free energies in TU-TILD simulations \cite{38}. As opposed to anharmonic excitations, FM effects tend to hinder the creation of Ni vacancies. For instance, they can increase $G^f$ by 0.03 eV in the ground state \cite{38}. If FM contributions are entirely removed, we will obtain an almost flawless agreement between the SMM and the TU-TILD \cite{38}. However, it should be emphasized that determining the magnetic properties of Ni is very challenging. In particular, correlations between the magnetic transition and the vacancy formation remain ambiguous due to the enormous computational expense of paramagnetic DFT calculations \cite{46,47,48}. More efforts are necessary to gain a comprehensive picture of Ni vacancies.  

\section{Conclusion}
We have modified the SMM to yield insight into the thermal generation of vacancies. Anharmonic effects have been appropriately included via high-order moments of the atomic displacement. On that basis, we have successfully explained non-Arrhenius behaviors of Ag, Cu, and Ni vacancies with a minimal computational workload. Our numerical analyses are quantitatively consistent with cutting-edge simulations and experiments. Therefore, our findings would be valuable for controlling the defect thermodynamics in FCC metallic systems. It is feasible to extend our quantum statistical model to other types of crystalline structures (see Electronic Supplementary Information). 

\section*{Acknowledgements}
T. D. Cuong is deeply grateful for the research assistantship at Phenikaa University. This research was funded by the Vietnam National Foundation for Science and Technology Development (NAFOSTED) under grant number 103.01-2019.318.

\end{document}


\title{Electronic Supplementary Information for\\ \textit{Theoretical Insights into Non-Arrhenius Behaviors of Thermal Vacancies in Anharmonic Crystals}}
\author{Tran Dinh Cuong}
\email{cuong.trandinh@phenikaa-uni.edu.vn}
\affiliation{Faculty of Materials Science and Engineering, Phenikaa University, Hanoi 12116, Vietnam}
\author{Anh D. Phan}
\email{anh.phanduc@phenikaa-uni.edu.vn}
\affiliation{Faculty of Materials Science and Engineering, Phenikaa University, Hanoi 12116, Vietnam}
\affiliation{Phenikaa Institute for Advanced Study (PIAS), Phenikaa University, Hanoi 12116, Vietnam}
\maketitle
\section{The heterogeneity of defective fcc crystals}
From the Leibfried-Ludwig theory [S1], we can encode structural information about the imperfect FCC system in 
\begin{gather}
k_i^V=\frac{1}{2}\left[\sum_{\substack{j\ne i\\j\ne l}}^{N}\frac{d^2\phi_{ij}}{d r_{ij}^2}\frac{r_{ij\alpha}^2}{r_{ij}^2}+\sum_{\substack{j\ne i\\j\ne l}}^{N}\frac{d\phi_{ij}}{d r_{ij}}\left(\frac{1}{r_{ij}}-\frac{r_{ij\alpha}^2}{r_{ij}^3}\right)-\frac{1}{2\rho_i^{1/2}}\sum_{\substack{j\ne i\\j\ne l}}^{N}\frac{d^2\varphi_{ij}}{d r_{ij}^2}\frac{r_{ij\alpha}^2}{r_{ij}^2}\right.\nonumber\\
\left.-\frac{1}{2\rho_i^{1/2}}\sum_{\substack{j\ne i\\j\ne l}}^{N}\frac{d\varphi_{ij}}{d r_{ij}}\left(\frac{1}{r_{ij}}-\frac{r_{ij\alpha}^2}{r_{ij}^3}\right)\textcolor{blue}{+\frac{1}{4\rho_i^{3/2}}\left(\sum_{\substack{j\ne i\\j\ne l}}^{N}\frac{d\varphi_{ij}}{d r_{ij}}\frac{r_{ij\alpha}}{r_{ij}}\right)^2}\right],
\label{S1}
\end{gather}
\begin{gather}
\gamma_{1i}^V=\frac{1}{48}\left\{\sum_{\substack{j\ne i\\j\ne l}}^{N}\frac{d^4\phi_{ij}}{d r_{ij}^4}\frac{r_{ij\alpha}^4}{r_{ij}^4}+\sum_{\substack{j\ne i\\j\ne l}}^{N}\frac{d^3\phi_{ij}}{d r_{ij}^3}\left(\frac{6r_{ij\alpha}^2}{r_{ij}^3}-\frac{6r_{ij\alpha}^4}{r_{ij}^5}\right)+\sum_{\substack{j\ne i\\j\ne l}}^{N}\frac{d^2\phi_{ij}}{d r_{ij}^2}\left(\frac{15r_{ij\alpha}^4}{r_{ij}^6}-\frac{18r_{ij\alpha}^2}{r_{ij}^4}+\frac{3}{r_{ij}^2}\right)\right.\nonumber\\
+\sum_{\substack{j\ne i\\j\ne l}}^{N}\frac{d\phi_{ij}}{d r_{ij}}\left(-\frac{15r_{ij\alpha}^4}{r_{ij}^7}+\frac{18r_{ij\alpha}^2}{r_{ij}^5}-\frac{3}{r_{ij}^3}\right)-\frac{1}{2\rho_i^{1/2}}\sum_{\substack{j\ne i\\j\ne l}}^{N}\frac{d^4\varphi_{ij}}{d r_{ij}^4}\frac{r_{ij\alpha}^4}{r_{ij}^4}-\frac{1}{2\rho_i^{1/2}}\sum_{\substack{j\ne i\\j\ne l}}^{N}\frac{d^3\varphi_{ij}}{d r_{ij}^3}\left(\frac{6r_{ij\alpha}^2}{r_{ij}^3}-\frac{6r_{ij\alpha}^4}{r_{ij}^5}\right)\nonumber\\
-\frac{1}{2\rho_i^{1/2}}\sum_{\substack{j\ne i\\j\ne l}}^{N}\frac{d^2\varphi_{ij}}{d r_{ij}^2}\left(\frac{15r_{ij\alpha}^4}{r_{ij}^6}-\frac{18r_{ij\alpha}^2}{r_{ij}^4}+\frac{3}{r_{ij}^2}\right)-\frac{1}{2\rho_i^{1/2}}\sum_{\substack{j\ne i\\j\ne l}}^{N}\frac{d\varphi_{ij}}{d r_{ij}}\left(-\frac{15r_{ij\alpha}^4}{r_{ij}^7}+\frac{18r_{ij\alpha}^2}{r_{ij}^5}-\frac{3}{r_{ij}^3}\right)\nonumber\\
+\frac{3}{4\rho_i^{3/2}}\left[\sum_{\substack{j\ne i\\j\ne l}}^{N}\frac{d^2\varphi_{ij}}{d r_{ij}^2}\frac{r_{ij\alpha}^2}{r_{ij}^2}+\sum_{\substack{j\ne i\\j\ne l}}^{N}\frac{d\varphi_{ij}}{d r_{ij}}\left(\frac{1}{r_{ij}}-\frac{r_{ij\alpha}^2}{r_{ij}^3}\right)\right]^2\nonumber\\
\textcolor{blue}{+\frac{1}{\rho_i^{3/2}}\left(\sum_{\substack{j\ne i\\j\ne l}}^{N}\frac{d\varphi_{ij}}{d r_{ij}}\frac{r_{ij\alpha}}{r_{ij}}\right)\left[\sum_{\substack{j\ne i\\j\ne l}}^{N}\frac{d^3\varphi_{ij}}{d r_{ij}^3}\frac{r_{ij\alpha}^3}{r_{ij}^3}+\sum_{\substack{j\ne i\\j\ne l}}^{N}\frac{d^2\varphi_{ij}}{d r_{ij}^2}\left(\frac{3r_{ij\alpha}}{r_{ij}^2}-\frac{3r_{ij\alpha}^3}{r_{ij}^4}\right)+\sum_{\substack{j\ne i\\j\ne l}}^{N}\frac{d\varphi_{ij}}{d r_{ij}}\left(-\frac{3r_{ij\alpha}}{r_{ij}^3}+\frac{3r_{ij\alpha}^3}{r_{ij}^5}\right)\right]}\nonumber\\
\left.\textcolor{blue}{-\frac{9}{4\rho_i^{5/2}}\left(\sum_{\substack{j\ne i\\j\ne l}}^{N}\frac{d\varphi_{ij}}{d r_{ij}}\frac{r_{ij\alpha}}{r_{ij}}\right)^2\left[\sum_{\substack{j\ne i\\j\ne l}}^{N}\frac{d^2\varphi_{ij}}{d r_{ij}^2}\frac{r_{ij\alpha}^2}{r_{ij}^2}+\sum_{\substack{j\ne i\\j\ne l}}^{N}\frac{d\varphi_{ij}}{d r_{ij}}\left(\frac{1}{r_{ij}}-\frac{r_{ij\alpha}^2}{r_{ij}^3}\right)\right]+\frac{15}{16\rho_i^{7/2}}\left(\sum_{\substack{j\ne i\\j\ne l}}^{N}\frac{d\varphi_{ij}}{d r_{ij}}\frac{r_{ij\alpha}}{r_{ij}}\right)^4}\right\},
\label{S2}
\end{gather}
\begin{gather}
\gamma_{2i}^V=\frac{1}{8}\left\{\sum_{\substack{j\ne i\\j\ne l}}^{N}\frac{d^4\phi_{ij}}{d r_{ij}^4}\frac{r_{ij\alpha}^2r_{ij\beta}^2}{r_{ij}^4}+\sum_{\substack{j\ne i\\j\ne l}}^{N}\frac{d^3\phi_{ij}}{d r_{ij}^3}\left(\frac{r_{ij\alpha}^2}{r_{ij}^3}+\frac{r_{ij\beta}^2}{r_{ij}^3}-\frac{6r_{ij\alpha}^2r_{ij\beta}^2}{r_{ij}^5}\right)+\sum_{\substack{j\ne i\\j\ne l}}^{N}\frac{d^2\phi_{ij}}{d r_{ij}^2}\left(-\frac{3r_{ij\alpha}^2}{r_{ij}^4}-\frac{3r_{ij\beta}^2}{r_{ij}^4}+\frac{15r_{ij\alpha}^2r_{ij\beta}^2}{r_{ij}^6}+\frac{1}{r_{ij}^2}\right)\right.\nonumber\\
+\sum_{\substack{j\ne i\\j\ne l}}^{N}\frac{d\phi_{ij}}{d r_{ij}}\left(\frac{3r_{ij\alpha}^2}{r_{ij}^5}+\frac{3r_{ij\beta}^2}{r_{ij}^5}-\frac{15r_{ij\alpha}^2r_{ij\beta}^2}{r_{ij}^7}-\frac{1}{r_{ij}^3}\right)-\frac{1}{2\rho_i^{1/2}}\sum_{\substack{j\ne i\\j\ne l}}^{N}\frac{d^4\varphi_{ij}}{d r_{ij}^4}\frac{r_{ij\alpha}^2r_{ij\beta}^2}{r_{ij}^4}-\frac{1}{2\rho_i^{1/2}}\sum_{\substack{j\ne i\\j\ne l}}^{N}\frac{d^3\varphi_{ij}}{d r_{ij}^3}\left(\frac{r_{ij\alpha}^2}{r_{ij}^3}+\frac{r_{ij\beta}^2}{r_{ij}^3}-\frac{6r_{ij\alpha}^2r_{ij\beta}^2}{r_{ij}^5}\right)\nonumber\\
-\frac{1}{2\rho_i^{1/2}}\sum_{\substack{j\ne i\\j\ne l}}^{N}\frac{d^2\varphi_{ij}}{d r_{ij}^2}\left(-\frac{3r_{ij\alpha}^2}{r_{ij}^4}-\frac{3r_{ij\beta}^2}{r_{ij}^4}+\frac{15r_{ij\alpha}^2r_{ij\beta}^2}{r_{ij}^6}+\frac{1}{r_{ij}^2}\right)-\frac{1}{2\rho_i^{1/2}}\sum_{\substack{j\ne i\\j\ne l}}^{N}\frac{d\varphi_{ij}}{d r_{ij}}\left(\frac{3r_{ij\alpha}^2}{r_{ij}^5}+\frac{3r_{ij\beta}^2}{r_{ij}^5}-\frac{15r_{ij\alpha}^2r_{ij\beta}^2}{r_{ij}^7}-\frac{1}{r_{ij}^3}\right)\nonumber\\
+\frac{1}{4\rho_i^{3/2}}\left[\sum_{\substack{j\ne i\\j\ne l}}^{N}\frac{d^2\varphi_{ij}}{d r_{ij}^2}\frac{r_{ij\alpha}^2}{r_{ij}^2}+\sum_{\substack{j\ne i\\j\ne l}}^{N}\frac{d\varphi_{ij}}{d r_{ij}}\left(\frac{1}{r_{ij}}-\frac{r_{ij\alpha}^2}{r_{ij}^3}\right)\right]\left[\sum_{\substack{j\ne i\\j\ne l}}^{N}\frac{d^2\varphi_{ij}}{d r_{ij}^2}\frac{r_{ij\beta}^2}{r_{ij}^2}+\sum_{\substack{j\ne i\\j\ne l}}^{N}\frac{d\varphi_{ij}}{d r_{ij}}\left(\frac{1}{r_{ij}}-\frac{r_{ij\beta}^2}{r_{ij}^3}\right)\right]\nonumber\\
\textcolor{blue}{+\frac{1}{2\rho_i^{3/2}}\left(\sum_{\substack{j\ne i\\j\ne l}}^{N}\frac{d\varphi_{ij}}{d r_{ij}}\frac{r_{ij\alpha}}{r_{ij}}\right)\left[\sum_{\substack{j\ne i\\j\ne l}}^{N}\frac{d^3\varphi_{ij}}{d r_{ij}^3}\frac{r_{ij\alpha}r_{ij\beta}^2}{r_{ij}^3}+\sum_{\substack{j\ne i\\j\ne l}}^{N}\frac{d^2\varphi_{ij}}{d r_{ij}^2}\left(\frac{r_{ij\alpha}}{r_{ij}^2}-\frac{3r_{ij\alpha}r_{ij\beta}^2}{r_{ij}^4}\right)+\sum_{\substack{j\ne i\\j\ne l}}^{N}\frac{d\varphi_{ij}}{d r_{ij}}\left(-\frac{r_{ij\alpha}}{r_{ij}^3}+\frac{3r_{ij\alpha}r_{ij\beta}^2}{r_{ij}^5}\right)\right]}\nonumber\\
\textcolor{blue}{+\frac{1}{2\rho_i^{3/2}}\left(\sum_{\substack{j\ne i\\j\ne l}}^{N}\frac{d\varphi_{ij}}{d r_{ij}}\frac{r_{ij\beta}}{r_{ij}}\right)\left[\sum_{\substack{j\ne i\\j\ne l}}^{N}\frac{d^3\varphi_{ij}}{d r_{ij}^3}\frac{r_{ij\alpha}^2r_{ij\beta}}{r_{ij}^3}+\sum_{\substack{j\ne i\\j\ne l}}^{N}\frac{d^2\varphi_{ij}}{d r_{ij}^2}\left(\frac{r_{ij\beta}}{r_{ij}^2}-\frac{3r_{ij\alpha}^2r_{ij\beta}}{r_{ij}^4}\right)+\sum_{\substack{j\ne i\\j\ne l}}^{N}\frac{d\varphi_{ij}}{d r_{ij}}\left(-\frac{r_{ij\beta}}{r_{ij}^3}+\frac{3r_{ij\alpha}^2r_{ij\beta}}{r_{ij}^5}\right)\right]}\nonumber\\
\textcolor{blue}{+\frac{1}{2\rho_i^{3/2}}\left(\sum_{\substack{j\ne i\\j\ne l}}^{N}\frac{d^2\varphi_{ij}}{d r_{ij}^2}\frac{r_{ij\alpha}r_{ij\beta}}{r_{ij}^2}-\sum_{\substack{j\ne i\\j\ne l}}^{N}\frac{d\varphi_{ij}}{d r_{ij}}\frac{r_{ij\alpha}r_{ij\beta}}{r_{ij}^3}\right)^2}\textcolor{blue}{-\frac{3}{8\rho_i^{5/2}}\left(\sum_{\substack{j\ne i\\j\ne l}}^{N}\frac{d\varphi_{ij}}{d r_{ij}}\frac{r_{ij\alpha}}{r_{ij}}\right)^2\left[\sum_{\substack{j\ne i\\j\ne l}}^{N}\frac{d^2\varphi_{ij}}{d r_{ij}^2}\frac{r_{ij\beta}^2}{r_{ij}^2}+\sum_{\substack{j\ne i\\j\ne l}}^{N}\frac{d\varphi_{ij}}{d r_{ij}}\left(\frac{1}{r_{ij}}-\frac{r_{ij\beta}^2}{r_{ij}^3}\right)\right]}\nonumber\\
\textcolor{blue}{-\frac{3}{8\rho_i^{5/2}}\left(\sum_{\substack{j\ne i\\j\ne l}}^{N}\frac{d\varphi_{ij}}{d r_{ij}}\frac{r_{ij\beta}}{r_{ij}}\right)^2\left[\sum_{\substack{j\ne i\\j\ne l}}^{N}\frac{d^2\varphi_{ij}}{d r_{ij}^2}\frac{r_{ij\alpha}^2}{r_{ij}^2}+\sum_{\substack{j\ne i\\j\ne l}}^{N}\frac{d\varphi_{ij}}{d r_{ij}}\left(\frac{1}{r_{ij}}-\frac{r_{ij\alpha}^2}{r_{ij}^3}\right)\right]}\nonumber\\
\textcolor{blue}{-\frac{3}{2\rho_i^{5/2}}\left(\sum_{\substack{j\ne i\\j\ne l}}^{N}\frac{d\varphi_{ij}}{d r_{ij}}\frac{r_{ij\alpha}}{r_{ij}}\right)\left(\sum_{\substack{j\ne i\\j\ne l}}^{N}\frac{d\varphi_{ij}}{d r_{ij}}\frac{r_{ij\beta}}{r_{ij}}\right)\left(\sum_{\substack{j\ne i\\j\ne l}}^{N}\frac{d^2\varphi_{ij}}{d r_{ij}^2}\frac{r_{ij\alpha}r_{ij\beta}}{r_{ij}^2}-\sum_{\substack{j\ne i\\j\ne l}}^{N}\frac{d\varphi_{ij}}{d r_{ij}}\frac{r_{ij\alpha}r_{ij\beta}}{r_{ij}^3}\right)}\nonumber\\
\left.\textcolor{blue}{+\frac{15}{16\rho_i^{7/2}}\left(\sum_{\substack{j\ne i\\j\ne l}}^{N}\frac{d\varphi_{ij}}{d r_{ij}}\frac{r_{ij\alpha}}{r_{ij}}\right)^2\left(\sum_{\substack{j\ne i\\j\ne l}}^{N}\frac{d\varphi_{ij}}{d r_{ij}}\frac{r_{ij\beta}}{r_{ij}}\right)^2}\right\},
\label{S3}
\end{gather}
\begin{gather}
\phi_{ij}=\varepsilon\left(\frac{a}{r_{ij}}\right)^n,\quad\varphi_{ij}=4\varepsilon^2c^2\left(\frac{a}{r_{ij}}\right)^m,\quad\rho_i=\sum_{\substack{j\ne i\\j\ne l}}^{N}\varphi_{ij}.
\label{S4}
\end{gather}
Here, blue terms arise from the breakdown of FCC symmetry. It is conspicuous that $k_i^V$, $\gamma_{1i}^V$, and $\gamma_{2i}^V$ depend on the relative position between the $i$th and $l$th lattice sites. Therefore, each atom has a different set of coupling parameters. This interesting characteristic cannot be found in perfect FCC structures.

To further clarify the heterogeneous nature, we apply Equations (S1)-(S4) to the 1st and 2nd neighbors of the point defect. As shown in Figure S1, while numerical results for 2nd neighbors are very close to ideal values, those for 1st ones are significantly lower. These discrepancies imply that the vacancy formation is dominated by the local dynamics. Our conclusion is consistent with recent DFT studies [S2,S3]. 

\begin{figure}[htp]
\renewcommand{\thefigure}{S\arabic{figure}}
\includegraphics[width=17 cm]{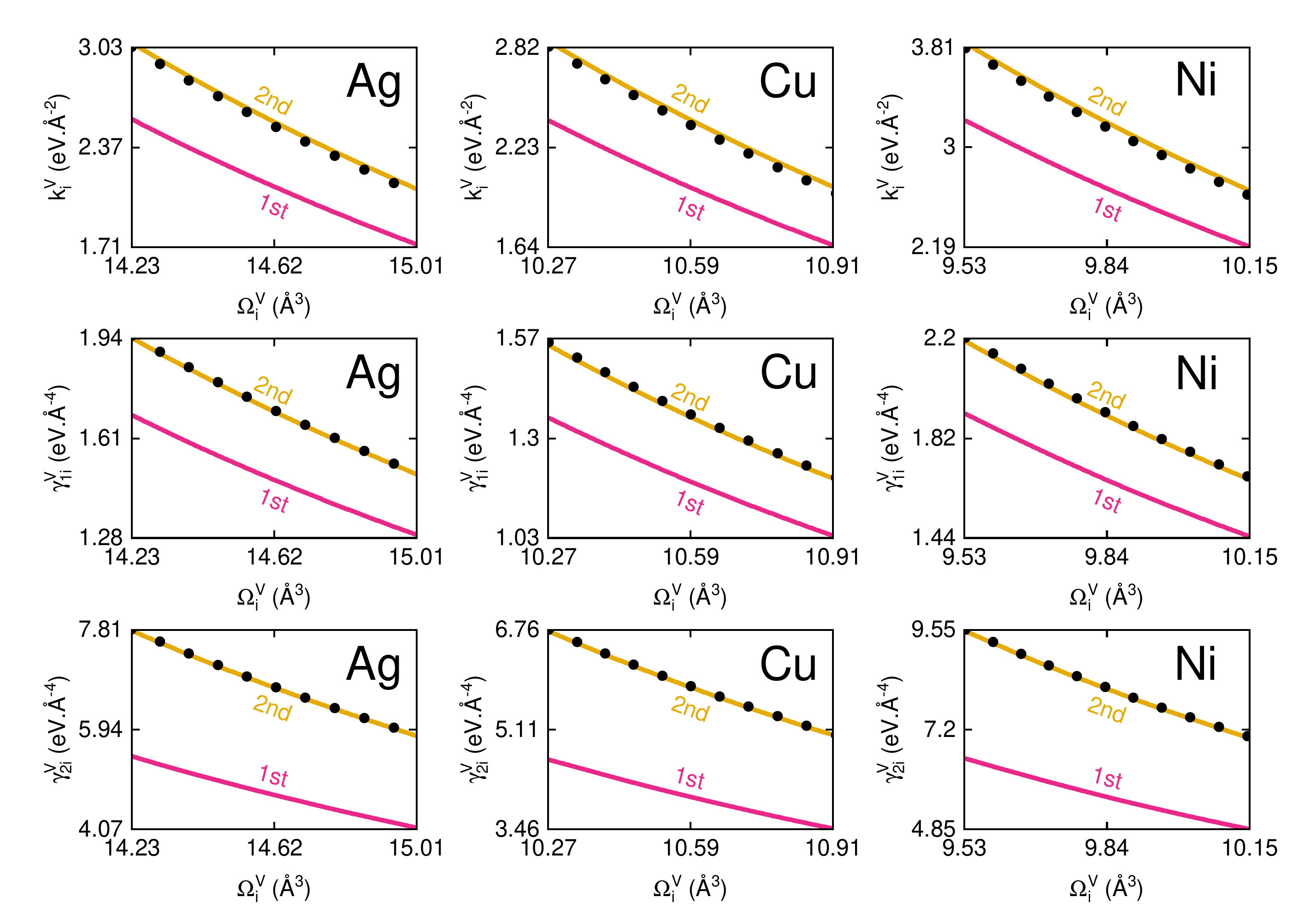}
\caption{(Color online) Coupling parameters versus atomic volumes in perfect (filled symbols) and imperfect (solid lines) FCC metals.}
\end{figure}

\section{The extension of SMM analyses for BCC systems}
According to Finnis and Sinclair [S4], the cohesion in BCC metals can be appropriately described by 
\begin{gather}
E_i=\sum_{j\ne i}^{N}\left(r_{ij}-c\right)^2\left(c_0+c_1r_{ij}+c_2r_{ij}^2\right)-2A\sqrt{\sum_{j\ne i}^{N}\left[\left(r_{ij}-d\right)^2+\frac{B}{d}\left(r_{ij}-d\right)^3\right]}.
\label{S5}    
\end{gather}
The first part of Equation (S5) denotes pairwise interactions, whereas the second part represents many-body effects. Finnis-Sinclair parameters ($c$, $c_0$, $c_1$, $c_2$, $A$, $d$, $B$) were determined in Ref.[S4] via experimental data for the lattice constant, the cohesive energy, and the elastic modulus. 

Based on the Finnis-Sinclair potential, we consider the thermal evolution of vacancies in two typical refractory materials, Mo and W. Remarkably, the loss of BCC symmetry leads to the emergence of a new coupling parameter, which is 
\begin{gather}
\beta_i^{V}=\frac{1}{2}\left(\frac{\partial^3E_i^V}{\partial u_{i\alpha}^V\partial u_{i\beta}^V\partial u_{i\eta}^V}\right)_{eq},
\label{S6}    
\end{gather}
where $u_{i\alpha}^V$ is the atomic displacement along the $\alpha$ axis ($\alpha\ne\beta\ne\eta=x,y,z$). Equation (S6) enables us to add odd-order contributions to the anharmonic free energy as [S5] 
\begin{gather}
F_{i}^{V(ANH)}=3\left(\int_0^{\gamma_{1i}^V}\langle u^{V4}_{i}\rangle d\gamma_{1i}^V+\int_0^{\gamma_{2i}^V}\langle u^{V2}_{i}\rangle^2 d\gamma_{2i}^V+\int_0^{\beta_i^V}\langle u^{V2}_{i}\rangle\langle u_{i}^V\rangle d\beta_i^V\right).
\label{S7}    
\end{gather}
Therefore, at high temperatures $\left(x_i\coth x_i\approx x_i^V\coth x_i^V\approx 1\right)$, the Gibbs energy of vacancy formation is given by 
\begin{eqnarray}
G^f&=&\sum_{i\ne l}^N\left\{\cfrac{1}{2}\left(E_i^V-E_i\right)+\cfrac{3}{2}\theta\ln\left(\cfrac{k_i^V}{k_i}\right)+3\theta^2\left[\cfrac{\gamma_{2i}^V-\gamma_{1i}^V}{K_i^{V2}}-\cfrac{\gamma_{2i}-\gamma_{1i}}{k_i^2}+\left(\frac{\gamma_i^V}{K_i^{V3}}\right)^{\frac{1}{2}}\frac{\beta_i^V}{K_i^V}\right]\right.\nonumber\\
&+&\left.3\theta^3\left[\cfrac{4\left(\gamma_{2i}^{V2}-3\gamma_{1i}^{V2}-6\gamma_{1i}^V\gamma_{2i}^V\right)}{K_i^{V4}}-\cfrac{4\left(\gamma_{2i}^2-3\gamma_{1i}^2-6\gamma_{1i}\gamma_{2i}\right)}{k_i^4}+\left(\frac{\gamma_i^V}{K_i^{V3}}\right)^{\frac{3}{2}}\beta_i^V+\frac{2k_i^V\gamma_i^V}{K_i^{V6}}\beta_i^{V2}\right]\right\},
\label{eq:S8}    
\end{eqnarray}
where $\gamma_i^V=4\left(\gamma_{1i}^V+\gamma_{2i}^V\right)$ and $K_i^V=k_i^V-\beta_i^{V2}/3\gamma_i^V$. As illustrated in Figure S2, the odd-order anharmonicity in defective BCC crystals is very large. This event naturally explains why $G^f$ plunges nonlinearly during heating. A strong correlation among asymmetric properties, anharmonic excitations, and non-Arrhenius behaviors is highlighted. Our SMM calculations agree quantitatively well with the latest atomistic simulations [S6,S7].

\begin{figure}[htp]
\renewcommand{\thefigure}{S\arabic{figure}}
\includegraphics[width=11.8 cm]{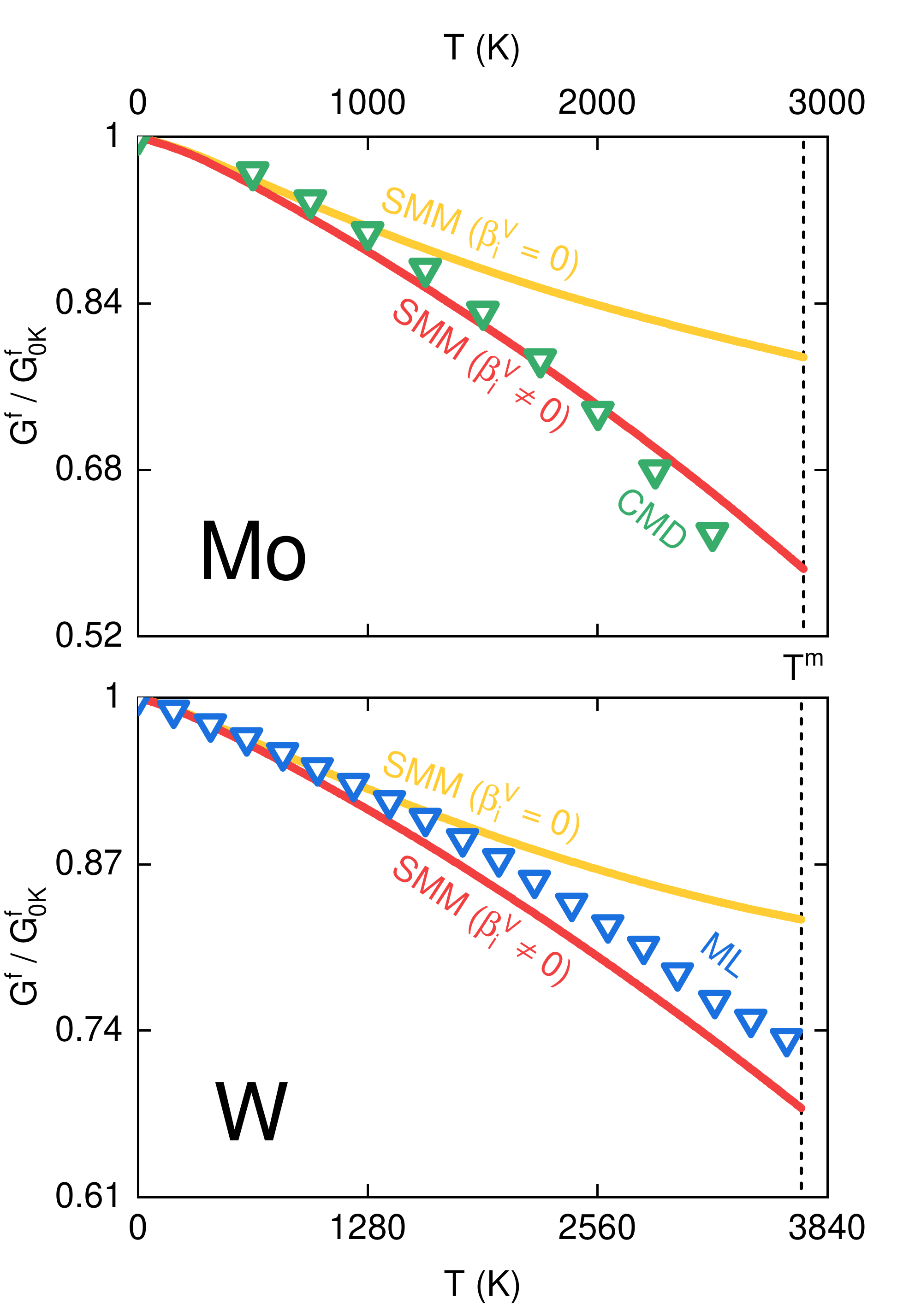}
\caption{(Color online) SMM analyses (solid lines) and CMD/ML computations [S6,S7] (open symbols) for the scaled Gibbs energy of vacancy formation in Mo and W.}
\end{figure}

